\PassOptionsToPackage{dvipsnames}{xcolor}	
\documentclass[12pt,letter]{article}

\usepackage[T1]{fontenc}
\usepackage{lmodern}
\usepackage{setspace}
\usepackage{etoolbox}
\makeatletter
\patchcmd{\appendix}{\@Alph}{\@Roman}{}{}
\makeatother

\usepackage{amsmath}
\usepackage{amssymb}
\usepackage{amsthm}
\usepackage{mathtools}
\usepackage{mathrsfs}
\usepackage{breqn}
\usepackage{bigints}

\usepackage[margin=1 in]{geometry}
\usepackage{enumerate}
\usepackage{enumitem} 
\usepackage{xcolor}
\usepackage[hidelinks]{hyperref}
\usepackage{float}
\usepackage{indentfirst}
\usepackage{caption}
\usepackage{fancyhdr}
\usepackage{tcolorbox}
\usepackage{multirow,array}
\usepackage{bbm}

\usepackage{tikz}
\usetikzlibrary{decorations.pathreplacing}
\usepackage{mathtools}
\usetikzlibrary{positioning}
\usepackage{graphicx}

\newcommand{\mcal}{\mathcal}

\renewcommand{\epsilon}{\varepsilon}

\newtheorem{theorem}{Theorem}

\newtheorem*{statement*}{Result}

\theoremstyle{definition}

\DeclareMathOperator*{\argmax}{\arg\!\max}
\DeclareMathOperator*{\argmin}{\arg\!\min}
\DeclareTextFontCommand{\emph}{\slshape}

\usepackage{natbib}
\nocite{*}

\title{Seeding an Uncertain Technology}

\author{Eric Gao\thanks{Department of Economics, MIT. ericgao@mit.edu. I would like to thank Mohammad Akbarpour, Joshua Gross, Matthew O. Jackson, Daniel Luo, and Eric Tang for helpful feedback.}}

\date{\today}

\begin{document}
	
\maketitle

\begin{abstract}
    I study how a startup with uncertainty over product quality and no knowledge of the underlying diffusion network optimally chooses initial seeds. To ensure widespread adoption when the product is good while minimizing negative perceptions when it is bad, the optimal number of initial seeds should grow logarithmically with network size. When there are agents of different types that govern their connectivity, it is asymptotically optimal to seed agents of a single type: the type that minimizes the marginal cost per probability of making the product go viral. These results rationalize startup behavior in practice.
\end{abstract}

	
\newpage

\onehalfspacing

\section{Introduction}

Many product launches start within a small, particular group: Facebook started at Harvard, Apple's first deal was for only 50 Apple I's with Byte Shop, Tesla's 2008 reveal was a 350-person event, and many video games have restrictive beta testing programs. One common explanation for such behavior is that new companies do not have the funding to engage in larger launches or are only familiar with a particular niche. However, it is always possible to secure additional funding or hire experts in other niches if small launches were not optimal. What, then, explains relatively contained product launches?
 
To answer this question, I formulate the problem a startup faces when deciding how to roll out a new product as an optimal seeding problem but with less initial information. Unlike established companies with more resources and expertise, startups face the unique challenge of having less information about how the product will spread. Such uncertainty comes from two potential sources. First, the startup may lack information about the quality of the product itself due to insufficient time and capital for extensive testing. Second, the startup may have little information on the underlying network structure that governs how adoption spreads. If maximizing adoption is the sole goal, one possible solution would be to simply seed everyone. However, this approach is clearly flawed: If the product is bad, everyone will have a bad impression of the product, leading to difficulties releasing a new and improved version of the product in the future.

Instead, the startup can choose some set of individuals to initially seed (i.e., beta testers). If the product is good, it will spread more widely; if bad, knowledge of its shortcomings remains localized. As such, exploiting network effects is one way to hedge against a potentially poor initial launch. This intuition guides the optimal seeding strategy. When the product is bad, components in the network are small and a constant size, leading to a constant ``marginal cost'' of additional seeds. When the product is good, a giant component emerges and additional seeding is most beneficial if it is the first seed to hit the giant component. As such, the ``marginal benefit'' of additional seeds decreases as the probability of the giant component already being hit grows.\footnote{For the purposes of this paper, all costs and benefits stem solely from the number of users. In the case of platforms such as Instagram and Facebook, this is indeed the case: Adding users is essentially costless and profits come from ads, etc. that scale with usage. However, the approach taken can easily be generalized to accommodate non-zero costs of seeding, which is discussed later.} Setting marginal cost equal to marginal benefit gives that the optimal number of seeds is logarithmic in network size. When there are many types of agents, it is furthermore optimal to only seed agents of a single type: the type that has the lowest ``marginal cost per probability'' of hitting the giant component when the product is good. These findings rationalize the observed behavior of many now-successful startups.

\paragraph{Related Literature} We relate most directly to the literature on optimal seeding. The closest paper in spirit is \cite{mcadams_2025_adoption}, which considers optimal marketing schemes when agents with private signals about product quality make strategic inferences about product quality and make an irreversible decision on whether to adopt the product. However, their results are driven by timing (if a product were good, agents would have heard about it sooner rather than later) instead of network structure.

Similarly, \cite{iyer_2019_when} consider when having too many initial seeds is potentially harmful. Their results stem from diffusion being driven by social forces: In their model, a customer's experience using the product depends on how many neighbors are also using the product. Early adopters may have initially bad experiences and stop using the product, even after more people join. They show, through simulations, that seeding the ``core'' of a network is often preferred to initial mass adoption.

My results are driven by inherent uncertainty in the quality of the product. Most prior literature studying optimal seeding assumes full knowledge of the diffusion process, so that there is no uncertainty in the technology or information being seeded. For example, consider \cite{sadler_2019_influence} which considers how to optimally influence public opinion in a network. However, there may be large uncertainties in how a particular message is perceived, especially varying based on ideological predispositions. A post highlighting one party's accomplishments may be subjectively interpreted by different individuals as:
\begin{itemize}
    \item A sign that that party can accomplish things.
    \item A sign indicating whether a party's priorities are right or wrong
    \item A joke or satirical if the divergence between beliefs is large enough.
\end{itemize}
Furthermore, these differences also influence how content propagates. After observing content, individuals may comment in support or opposition of the original post; they may also re-post, share, or ``quote'' the post, adding their own thoughts, whether critical or supportive. Experimental testing to form precise beliefs about a post's performance is infeasible due to the fast pace of politics.

 \cite{akbarpour_2023_just} also considers an inhomogeneous random network model to study the value of optimal seeding versus simply introducing additional seeds. They find that whenever significant diffusion is feasible, random seeding with a few more seeds outperforms optimal seeding. However, the sole objective in this setting is to maximize diffusion---taking this logic to its limit, the optimal policy is to just seed everyone. Yet in the context of information dissemination, \cite{banerjee_2023_when} finds that \textit{selective} seeding outperforms mass seeding, albeit for different reasons than the forces considered in this paper. In their setting, mass seeding (and in particular, common information of the fact that everyone is initially exposed to the information) discourages individuals from asking clarifying questions out of fear of seeming slow to process information. 

Other papers have also studied the optimal seeding problem. \cite{banerjee_2019_using} finds that individuals within a particular network are good at identifying influential nodes (without macro-level knowledge of the entire network structure). \cite{sadler_2025_seeding} derives tractable and less measurement-intensive methods to compute which group of individuals within a network should be seeded. \cite{keng_2025_contagion} studies the (random) linear threshold model where one node adopts a technology if at least a constant fraction of their neighbors have already adopted, and characterizes the probability of full contagion based on the set of initial adopters. Unfortunately, a common theme in the literature is that analytically finding the optimal (set of) seeds is generally intractable.

\section{Model}

We use the inhomogeneous random networks (IRN) model developed by \cite{bollobs_2007_the}. A network is a graph $G = (N, E)$ consisting of a set of agents $N$ and a set of edges $E \subset N \times N$. We consider sequences of networks as $|N|$ grows large, as is common in the literature. Each agent has some type $n_i$ drawn from the set $\mcal T$, with $|N_i|$ denoting the number of agents of type $i$. For each $i$, let 
$$\mu(i) = \lim_{|N| \to \infty} \frac{|N_i|}{|N|}$$
represent the proportion of agents that are of type $i$ when the network grows large. There are two states of the world: Either the product is good ($G$) or it is bad ($B$).\footnote{A richer model in which the designer has a distribution of beliefs over more than two states of the world, with probabilities of users having good or bad experiences depending on the state, exhibits similar qualitative behavior but is much less tractable.}

An agent of type $i$ shares an edge with each agent of type $j$ with probability $\kappa^X(i, j)/|N|$ when the state of the world is $X \in \{G, B\}$. As the network is undirected, $\kappa^X(i,j) = \kappa^X(j, i)$ for all $i, j \in \mcal T$. For large networks, $\kappa^X(i, j)$ is then the expected number of type $j$ neighbors an agent of type $i$ has when the state of the world is $X$. Let
$$T^X = \left[\kappa^X(i, j)\right]_{i, j \in \mcal T} \text{ for } X \in \{G, B\}$$
denote the kernel matrix when the state of the world is $X$. The largest eigenvalue of $T^X$, denoted $\lambda_1^X$, determines the macroscopic behavior of the IRN. If this eigenvalue is less than one, there are many small disjoint components with the largest having $O(\log |N|)$ agents. If the largest eigenvalue is more than one, then there is one giant component containing a positive fraction of all the agents (and many smaller disjoint components once again bounded by $O(\log |N|)$ in size). We assume that $\lambda_1^G > 1$ and $\lambda_1^B < 1$ to keep the problem interesting. Economically, this corresponds to good products blowing up and becoming viral and bad products failing to do so as individuals are unlikely to adopt bad products. 

The designer knows agent types but not the underlying network structure and chooses to seed a subset of agents $S \subseteq N$. This is equivalent to the designer choosing a number of initial seeds of each type and then randomly seeding accordingly, as the designer does not know the realized network. In many social media platform examples, the set of types can be thought of as which university an individual attends, with the designer being able to choose how initial seeds are dispersed or concentrated across universities (e.g., Facebook, BeReal, and Fizz all initially started on college campuses). After initial seeds are chosen, nature draws the state of the world from $\{G, B\}$ and a network is realized according to either $T^G$ or $T^B$. An agent adopts the technology if there is a path from some initial seed to them (i.e. agent $i$ adopts if they are in the same component as an initially seeded agent $j$). Let $A^X(N, S)$ denote the expected number of agents who adopt when the state of the world is $X$, the set of agents is $N$, and the set of initial seeds is $S$. The designer values adoption when the state of the world is good and dislikes adoption when the state of the world is bad. Let
$$S^*(N) = \argmax_{S \subseteq N} \left\{ A^G(S, N) - \lambda A^B(S, N) \right\}$$
be the solution to the designer's problem, where $\lambda$ encodes the designer's relative weights for each state of the world. For example, $\lambda$ can incorporate the designer's prior belief over the two states or whether the designer cares more or less about good versus bad customer experiences. We seek to characterize $ S^*(N)$ as $|N|$ grows large.

\section{Optimal Seeding}

Optimal seeding takes on a simple structure: Seed an ``optimal'' type a logarithmic number of times, where ``optimal'' can be determined purely by model fundamentals. 
\begin{theorem}
    For any kernels $\kappa^G, \kappa^B$:
    \begin{enumerate}
        \item $|S^*(N)| = \Theta(\log |N|)$;
        \item It is asymptotically optimal (at rate $O(1/|N|)$ to only seed agents of the type which has the lowest marginal cost per probability of reaching the giant component. 
    \end{enumerate} 
\end{theorem}

We will prove the first point in the simpler case of an Erd\H{o}s-R\'enyi random graph (a special case of an IRN when there is only a single type). Next, we will take a more reduced-form approach to prove the second point. Intuitions will be developed along the way.

\begin{proof}
    Consider an Erd\H{o}s-R\'enyi random graph with one type $i$ and let $\kappa^X = \kappa^X(i, i)$ be the expected number of neighbors each agent has. In this setting, the designer simply chooses the number of agents to initially seed. Our assumption about subcritical behavior when the product is bad and supercritical behavior when the product is good is then $\kappa^B < 1, \kappa^G > 1$.

    When the product is bad, components have expected size $\frac{1}{1-\kappa^B}$ and with probability one as $|N|$ grows large, the largest component is $O(\log |N|)$. As such, every component is vanishing relative to the network. This implies that as long as
    $$\lim_{|N| \to \infty} \frac{\log(|N|) \cdot |S^*(N)|}{|N|} = 0$$
    the probability of seeding any component twice goes to zero; note that this is satisfied by $|S^*(N)| = \Theta(\log |N|)$. Thus,
    $$A^B(S, N) = |S| \cdot \frac{1}{1-\kappa^B}.$$

    When the product is good, there is a giant component that contains a positive fraction of all agents. In particular, let $y$ be the solution to 
    $$1-y = \exp\left({-\kappa^G y}\right).$$
    Then, with probability one as $|N|$ grows large, there is a component of size $y|N|$ and each agent has a probability of $y$ of being in the giant component. Each remaining component has expected size $\frac{1}{1-(1-y)\kappa^G}$. What is $A^G(S, N)$? The chance that the giant component is not seeded is $(1-y)^{|S|}$ and in expectation, $|S|(1-y)$ seeds do not hit the giant component. Thus, 
    $$A^G(S, N) = \left(1-(1-y)^{|S|}\right) y|N| + |S|(1-y) \cdot \frac{1}{1-(1-y)\kappa^G}.$$

    With $A^B$ and $A^G$ in mind, we now solve the designer's problem. The ``marginal cost'' of another seed is
    $$\frac{\lambda}{1-\kappa^B}$$
    as each additional seed hits a component of that size in expectation, scaled by $\lambda$. Note that this marginal cost is constant: It does not depend on which agents the designer has already seeded. The ``marginal benefit'' of another seed when the designer has already seeded $|S|$ agents is
    $$y \cdot \left[(1-y)^{|S|} y|N| \right]+ (1-y) \left[\frac{1}{1-(1-y)\kappa^G}\right].$$
    Intuitively, with probability $y$ the additional seed hits the giant component but is useful only if the giant component has not already been seeded. Then, the designer seeds as long as the marginal benefit outweighs the marginal cost:
    $$y \cdot \left[(1-y)^{|S|} y|N| \right]+ (1-y) \left[\frac{1}{1-(1-y)\kappa^G}\right] \geq \frac{\lambda}{1-\kappa^B}.$$
    Solving for $|S|$ gives
    $$|S| \leq \frac{1}{\log\left(\frac{1}{1-y}\right)} \left[\log(|N|) + 2 \log(y) - \log\left( \frac{\lambda}{1-\kappa^B} - \frac{1-y}{1-(1-y)\kappa^G}\right)\right]$$
    so
    $$|S^*(N)| = \left \lceil \frac{1}{\log\left(\frac{1}{1-y}\right)} \left[\log(|N|) + 2 \log(y) - \log\left( \frac{\lambda}{1-\kappa^B} - \frac{1-y}{1-(1-y)\kappa^G}\right)\right]\right \rceil.$$
    As $y, \lambda, \kappa^B, \kappa^G$ are constants, $|S^*(N)| = O(\log |N|)$. In the limit as the graph grows large, only $y$ matters; this corresponds to all other parameters only affecting a vanishing portion of the network. 
    
    To illustrate the solution, consider the problem Instagram faced when it launched in 2010. Back then, the world population was around 7 billion, which we use for $|N|$. We can estimate $y$ by taking the current fraction of people who use Instagram: approximately 2 billion out of the world's 8.1 billion. Plugging these estimates into the non-vanishing portion of $|S^*(N)|$ gives
    $$|S^*(N)| = \frac{\log(|N|)}{\log\left(\frac{1}{1-y}\right)} \approx \frac{\log(7,000,000,000)}{\log\left(\frac{1}{1-2/8.1}\right)} \approx 80$$
    initial seeds to be optimal. Instagram co-founder Kevin Systrom once stated that Instagram indeed started with around 100 beta testers who spread the app before it officially launched.\footnote{\url{https://www.quora.com/How-many-beta-users-did-Instagram-have-right-before-launch}} While this back-of-the-envelope calculation and model is not a perfect representation of the world, our results align well with observed data.

    Let us now consider the IRN case. Behavior is similar, but component sizes are more difficult to pin down. In this model, each node of type $i$ has probability $y(i)$ of being in the giant component when the product is good, where the function $y(\cdot)$ solves
    $$1-y(i) = \exp\left({-\int_{j \in \mcal T} \kappa^G(i, j)y(j)d\mu(j)}\right) \text{ for all } i \in \mcal T.$$
    Then, the expected size of the giant component when the product is good is
    $$y|N| \text{ for } y = \int_{j \in \mcal T} y(j)d\mu(j).$$
    The expected size of the small component a type $i$ agent is in when the product is bad is
    $$C^B(i) = [(I - T^B)^{-1} \mathbbm{1}]_i$$
    where the subscript $i$ denotes the $i$th component of a vector. Similarly, the expected size of the small component a type $i$ agent is in when the product is good, conditional on not being part of the giant component, is
    $$C^G(i) = [(I - \hat{T}^G)^{-1} \mathbbm{1}]_i$$
    where
    $$\hat{T}^G = \left[\kappa^G(i,j) (1-y(j))\right]_{i, \in \mcal Tj}$$
    is the dual kernel governing how agents not in the giant component are connected.

    In the Erd\H{o}s-R\'enyi random graph case, there was only one decision: when to stop seeding. In the general IRN case, the problem of selecting which types to seed is a combinatorially difficult integer problem. Instead, we will first consider a continuous relaxation by dropping the requirement that the number of seeds of each type must be an integer. Repeating a similar process as before, the marginal utility (marginal benefit minus marginal cost) from seeding an agent of type $i$ when the probability of seeding the giant component is $q$ is
    $$y(i) \cdot \left[(1-q) y|N|\right)] + (1-y(i)) C^G(i) - \lambda C^B(i).$$
    Then, a seed of type $i$ should be seeded until 
    $$q \geq 1- \frac{\lambda C^B(i) - (1-y(i)) C^G(i)}{y(i)y|N|}$$
    so the optimal probability to seed the giant component must be
    $$q^*(N) = \max_{i \in \mcal T}\left\{1- \frac{\lambda C^B(i) - (1-y(i)) C^G(i)}{y(i)y|N|}\right\}$$
    since any less would mean the marginal utility of seeding some type is strictly positive, and any more would mean that a marginal reduction in seeds of any type would be strictly beneficial. With this optimal probability fixed, the optimal set of seeds must then solve the following optimization problem:
    \begin{equation}\label{opt1}
        \begin{split}
            S^R(N) = \argmin_{\{S_i\}_{i \in \mcal T}} & \left\{ \sum_{i \in \mcal T} S_i \left[\lambda C^B(n_i) - (1-y(n_i)) C^G(n_i)\right] \right\}\\
        \text{s.t.} & 1-\prod_{i \in \mcal T} (1-y(i))^{S_i} \geq q^*(N)
        \end{split}
    \end{equation}
    where $S^R(N)$ specifies a (possibly non-integer) number $S_i$ of seeds of type $i$ to seed under the relaxed problem. Taking logarithms of the constraint gives (\ref{opt1}) to be equivalent to 
    \begin{equation}\label{opt2}
        \begin{split}
            S^R(N) = \argmin_{S \subset N} & \left\{ \sum_{i \in \mcal T} S_i \left[\lambda C^B(n_i) - (1-y(n_i)) C^G(n_i)\right] \right\}\\
        \text{s.t.} & \sum_{i \in \mcal T} \left[S_i\log(1-y(i))\right] \leq \log(1-q^*(N)).
        \end{split}
    \end{equation}
    This is now a linear program with a linear constraint, so the solution is to only seed agents of type
    \begin{equation*}
        j^* = \argmin_{j \in \mcal T}\left\{\frac{\lambda C^B(j) - (1-y(j)) C^G(j)}{-\log(1-y(j))}\right\}
    \end{equation*}
    until the probability of seeding the giant component hits at least $q^*(N)$. In particular, this corresponds to seeding
    $$S^R_{j^*}(N) = \frac{\log(1-q^*(N))}{\log(1-y(j))}.$$
    Finally, undoing the relaxation by rounding up gives
    $$S^*_{j^*}(N) = \left \lceil S^R_{j^*}(N) \right \rceil.$$
    The designer's utility under $S^*_{j^*}(N)$ is at most $\lambda C^B(j^*) - (1-y(j^*)) C^G(j^*)$ less than the designer's utility under $S^R_{j^*}(N)$ in the relaxed problem. The designer's utility under the relaxed problem is an upper bound of the designer's utility under the optimal solution to the original integer problem. Then, the designer's value from the problem grows as 
    \begin{align*}
        \Theta(q^*(N)\cdot|N|) & = \Theta\left(\max_{i \in \mcal T}\left\{1- \frac{\lambda C^B(i) - (1-y(i)) C^G(i)}{y(i)y|N|}\right\}|N|\right) \\
        & = \Theta\left(\max_{i \in \mcal T}\left\{|N|- \frac{\lambda C^B(i) - (1-y(i)) C^G(i)}{y(i)y}\right\}\right) = \Theta(|N|)
    \end{align*}
    so the constant gap between seeding only $j^*$ and the optimal solution vanishes at rate $O(1/|N|)$.

\end{proof}

Empirically, this parallels how many startups have initial users from a very specific group: Facebook initially targeted Harvard students, WhatsApp was first used by Russian immigrants in the Bay Area, and Spotify's beta testers were Swedish music bloggers. In the case of WhatsApp, the initial app was poorly received and only gained traction after integrating new smartphone updates into their product, which makes minimizing negative initial impressions important. These solutions also require minimal knowledge to implement: The startup only needs to know marginal costs/benefits and probabilities for type $j^*$.

\section{Discussion}

I view these results as complementary to other forces that may explain how startups behave when faced with the optimal seeding problem. Conventional wisdom suggests starting in a specific niche to better know an audience or focusing on one geographic area to better understand regulatory red tape. My model can be easily adapted to accommodate both situations. A better understanding of certain audiences can be modeled via changes in $\kappa^G, \kappa^B$, leading certain types to have different $y(i), C^G(i), C^B(i)$ values. Similarly, adding constant marginal costs to seeding individual agents does not change the logarithmic scaling of optimal seeding while costs based on the number of types seeded further pushes the designer to seeding only a single group. My results hold under both of these generalizations.

Another potential concern is that the IRN model is not representative of real-world networks. In particular, many social media apps are used only if there is high clustering among friends, but clustering is generally not present in random graphs. This provides another rationale for keeping initial seeds centralized and is especially pertinent in cases such as Facebook, where a critical mass of individuals needed to be seeded for anything to happen. My work provides an intuitive answer to the question of why Facebook didn't seed a second or third critical mass of early adopters at universities outside of Harvard. If Facebook was a good product then a single mass of early adopters is enough to make the product go viral with high probability; otherwise, those additional masses would have only led to additional bad impressions if the product was bad.

My work also contributes to the ex-ante design of optimal seeding strategies. In the standard influence maximization problem, the designer has some exogenously determined number of seeds and richer network information and needs to determine which particular agents (instead of just agent types) are optimal to seed. However, upstream to this problem, the designer often needs to acquire funding (a startup pitching to VC firms before having the resources to conduct market research, an experimental economist writing a grant before conducting a field experiment) before knowing network structure; this initial amount of funding then becomes the constraint on the number of seeds in the downstream influence maximization problem. Combined with \cite{akbarpour_2023_just}'s insight about how a constant number of additional seeds is more impactful than rich network knowledge, the designer's initial funding application should grow logarithmically in the size of the network.

There are several avenues for further research. Most interesting is incorporating dynamics into this model of diffusion. Beta testers are not only useful for spreading the word, but also for providing feedback about the product. Similarly, additional users also provide the designer with revenue that can be re-invested in research and development to improve the product. As such, product quality endogenously changes as more and more agents adopt the product. With a better-quality product, the designer may want to seed new agents, which matches waves of ``brand ambassadors'' for various companies being observed in the world. Designers may also have preferences for the speed of adoption, which a dynamic model would also address.

\newpage

\bibliography{cites}

@article{banerjee_2021_changes,
  author = {Banerjee, Abhijit V. and Breza, Emily and Chandrasekhar, Arun G. and Duflo, Esther and Jackson, Matthew O. and Kinnan, Cynthia},
  title = {Changes in Social Network Structure in Response to Exposure to Formal Credit Markets},
  doi = {10.2139/ssrn.3772604},
  urldate = {2021-12-29},
  year = {2021},
  journal = {SSRN Electronic Journal}
}

@article{banerjee_2013_the,
  author = {Banerjee, A. and Chandrasekhar, A. G. and Duflo, E. and Jackson, M. O.},
  month = {07},
  pages = {1236498-1236498},
  title = {The Diffusion of Microfinance},
  doi = {10.1126/science.1236498},
  urldate = {2019-12-20},
  volume = {341},
  year = {2013},
  journal = {Science}
}

@article{banerjee_2019_using,
  author = {Banerjee, Abhijit and Chandrasekhar, Arun G and Duflo, Esther and Jackson, Matthew O},
  month = {02},
  pages = {2453-2490},
  title = {Using Gossips to Spread Information: Theory and Evidence from Two Randomized Controlled Trials},
  doi = {10.1093/restud/rdz008},
  urldate = {2020-03-24},
  volume = {86},
  year = {2019},
  journal = {The Review of Economic Studies}
}

@article{banerjee_2023_when,
  author = {Banerjee, Abhijit and Breza, Emily and Chandrasekhar, Arun G and Golub, Benjamin},
  month = {07},
  pages = {1884-1922},
  publisher = {Oxford University Press},
  title = {When Less Is More: Experimental Evidence on Information Delivery During India’s Demonetisation},
  doi = {10.1093/restud/rdad068},
  urldate = {2025-01-30},
  volume = {91},
  year = {2023},
  journal = {The Review of Economic Studies}
}

@article{akbarpour_2023_just,
  author = {Akbarpour, Mohammad and Malladi, Suraj and Saberi, Amin},
  title = {Just a Few Seeds More: The Inflated Value of Network Data for Diffusion},
  url = {https://web.stanford.edu/~mohamwad/NetworkSeeding.pdf},
  urldate = {2025-03-16},
  year = {2023},
  journal = {}
}

@article{sadler_2025_seeding,
  author = {Sadler, Evan},
  month = {01},
  pages = {71-93},
  publisher = {Wiley},
  title = {Seeding a Simple Contagion},
  doi = {10.3982/ecta22448},
  url = {https://ideas.repec.org/a/wly/emetrp/v93y2025i1p71-93.html},
  urldate = {2025-03-09},
  volume = {93},
  year = {2025},
  journal = {Econometrica}
}

@article{keng_2025_contagion,
  author = {Keng, Ying Ying and Kwa, Kiam Heong},
  month = {02},
  pages = {129090},
  title = {Contagion probability in linear threshold model},
  doi = {10.1016/j.amc.2024.129090},
  urldate = {2024-11-02},
  volume = {487},
  year = {2025},
  journal = {Applied Mathematics and Computation}
}

@misc{harris_2015_direct,
  author = {Harris, Kameron and Payne, Joshua and Dodds, Peter},
  title = {Direct computation of contagion triggering probabilities for generalized and bipartite random networks},
  url = {https://pdodds.w3.uvm.edu/research/papers/years/2015/harris2015a.pdf},
  urldate = {2025-03-16},
  year = {2015}
}

@article{sadler_2019_influence,
  author = {Sadler, Evan},
  title = {Influence Campaigns},
  doi = {10.2139/ssrn.3371835},
  urldate = {2019-09-05},
  year = {2019},
  journal = {SSRN Electronic Journal}
}

@article{bollobs_2007_the,
  author = {Bollobás, Béla and Janson, Svante and Riordan, Oliver},
  pages = {3-122},
  title = {The phase transition in inhomogeneous random graphs},
  doi = {10.1002/rsa.20168},
  urldate = {2022-12-28},
  volume = {31},
  year = {2007},
  journal = {Random Structures and Algorithms}
}

@article{iyer_2019_when,
  author = {Iyer, Shankar and Adamic, Lada A.},
  month = {06},
  title = {When can overambitious seeding cost you?},
  doi = {10.1007/s41109-019-0146-z},
  urldate = {2025-02-23},
  volume = {4},
  year = {2019},
  journal = {Applied Network Science}
}

@article{mcadams_2025_adoption,
  author = {McAdams, David and Song, Yangbo},
  month = {01},
  pages = {453-480},
  publisher = {Econometric Society},
  title = {Adoption epidemics and viral marketing},
  doi = {10.3982/te5886},
  urldate = {2025-06-04},
  volume = {20},
  year = {2025},
  journal = {Theoretical Economics}
}

\end{document}